\documentclass[a4paper,11pt]{article}
\usepackage{pos}

\usepackage[utf8]{inputenc}
\usepackage[T1]{fontenc}
\usepackage{xcolor}
\usepackage{graphicx}
\usepackage{amsmath}
\usepackage{wrapfig}
\usepackage{amssymb}
\usepackage{subfigure}
\usepackage{siunitx}
\DeclareSIUnit{\ev}{eV}

\title{Seasonal Variations of the Unfolded Atmospheric Neutrino Spectrum with IceCube}

\ShortTitle{Seasonal Variations of the Atmospheric Neutrino Spectrum}
\author{The IceCube Collaboration \\{\normalsize \normalfont(a complete list of authors can be found at the end of the proceedings)}}

\emailAdd{karolin.hymon@tu-dortmund.de}
\emailAdd{tim.ruhe@tu-dortmund.de}

\abstract{The IceCube Neutrino Observatory is a detector array at the South Pole with the central aim of studying astrophysical neutrinos. However, the majority of the detected neutrinos originates from cosmic ray interactions in the atmosphere. The rate of these atmospheric neutrinos shows a seasonal variation indicating that the rate changes with the temperature in the stratosphere. These seasonal changes of the atmospheric neutrino energy spectrum will be investigated using the Dortmund Spectrum Estimation Algorithm (DSEA). Based on results obtained from 10\% of IceCube's atmospheric muon neutrino data, taken between 2011 and 2018, the differences of the measured fluxes during the Austral summer and winter will be discussed.

\vspace{4mm}
{\bfseries Corresponding authors:}
Karolin Hymon$^{1*}$, Tim Ruhe$^{1}$\\
{$^{1}$ \itshape Department of Physics, TU Dortmund University, 44221 Dortmund, Germany}\\
$^*$ Presenter

\FullConference{37$^{\rm{th}}$ International Cosmic Ray Conference (ICRC 2021)\\
		July 12th -- 23rd, 2021\\
		Online -- Berlin, Germany}

}

\begin{document}
\maketitle

\section{Introduction}

The atmospheric neutrino spectrum imposes, additionally to cosmic air shower physics, the main background in the measurement of astrophysical neutrinos.  Since the observed rate of atmospheric neutrinos correlates with the atmospheric temperature \cite{desiati11,Gaisser13,Heix19}, a   seasonal variation of the energy spectrum is expected to occur at energies above approximately  \SI{100}{\giga\ev} \cite{honda15}. 
In this contribution, we present an analysis of the atmospheric muon neutrino spectrum using unfolding techniques to investigate the detection of seasonal variations from 8 years of IceCube data.  


IceCube is a cubic-kilometer neutrino detector installed in the ice at the geographic South Pole between depths of \SI{1450}{\metre} and \SI{2450}{\metre}, completed in 2010. Reconstruction of the direction, energy and flavor of the neutrinos relies on the optical detection of Cherenkov radiation emitted by charged particles produced in the interactions of neutrinos in the surrounding ice or the nearby bedrock \cite{Aartsen:2016}. 

The majority of the detected neutrinos originates from meson decays in the muonic component of cosmic ray air showers. 
Whether these mesons decay directly into muons and neutrinos, or produce  secondary mesons,  depends on the local air density of  the atmosphere \cite{Gaisser13}. 
The resulting muon neutrino ($\nu_{\mu}$) and anti-neutrino ($\bar \nu_{\mu}$) flux  is given by the integration of the neutrino production yield 
over the slant depth $X$,

\begin{equation}
 \Phi_{\nu}(E_{\nu},\Theta,X) = \Phi_{\mathrm{N}}(E_{\nu}) \cdot
 \int\limits_0^{X_{\mathrm{ground}}}
 \left( \frac{A_{\pi\rightarrow\nu}(X)}{1 + B_{\pi\rightarrow\nu}(X) \cdot \frac{E_{\nu}\cos(\Theta^{\star})}{\epsilon_{\pi}(T(X))}} + \frac{A_{K\rightarrow\nu}(X)}{1 + B_{K\rightarrow\nu}(X) \cdot \frac{E_{\nu}\cos(\Theta^{\star})}{\epsilon_{K}(T(X))}} \right)  \mathrm{d}X, \label{neutrinoflux} 
\end{equation}
with the primary cosmic ray flux $\Phi_{\mathrm{N}}(E_{\nu})$ of nucleon N at neutrino energy $E_{\nu}$ \cite{gaisser90}; the quantity $A_{i\rightarrow\nu}$ accounts for decay branching ratios,  $B_{i\rightarrow\nu}$  for the cross sections; the denominator characterizes the competition between kaon and pion decays and the production of secondary mesons. 

The latter process is favored at critical energies $\epsilon_i$ above $E_{\nu} \cdot \cos(\Theta^{\star})$  at the zenith angle of neutrino production $\Theta^{\star}$. In this scenario the neutrino spectrum follows a steep power law $E^{-\gamma}$ with a spectral index of approximately $\gamma \approx 3.7$ \cite{Gaisser13}. Since the critical energy at a given atmospheric depth is anti-proportional to the air density, it is affected by temperature changes \cite{gaisser90}. Hence, the temperature becomes linearly correlated to the critical energy assuming the atmospheric isothermal approximation of the ideal gas law. The critical energies for kaons and pions are  $\epsilon_{\pi}\approx \SI{125}{\giga\ev}$ and $\epsilon_{K}\approx \SI{850}{\giga\ev}$ \cite{gaisser90}.

\begin{wrapfigure}{r}{0.5\textwidth}
\vspace{-20pt}
  \begin{center}
    \includegraphics[width=0.35\textwidth]{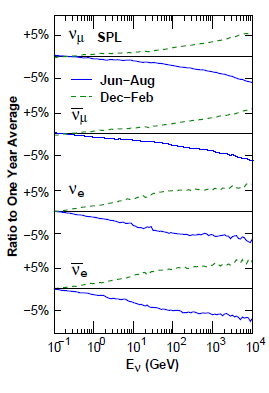}
 \end{center}
  \vspace{-10pt}
  \caption[Expected seasonal neutrino rate at South Pole ]{Ratio of the calculated neutrino flux for the  Austral summer and winter at South Pole for the zenith region between \SIrange{90}{120}{\degree} compared to the yearly average \cite{honda15}. 
   \label{honda_plot}}
  \vspace{5pt}
\end{wrapfigure}

The neutrino flux at South Pole based on the  NRLMSISE-00 atmospheric model \cite{nrlm} is displayed in Fig.\ \ref{honda_plot} \cite{honda15}. As expected, a distinct variation can be observed above the respective critical energies, further increasing with the neutrino energy.

\section{Spectrum Unfolding}


The reconstruction of the neutrino energy spectrum relies on the indirect measurement of neutrino-induced muons by the charged-current interaction inside the ice or the bedrock  \cite{ic59}. However, these 
muons are exposed to stochastic energy losses during their propagation through the  detector  \cite{etrun13} and in addition, the energy reconstruction is smeared by the limited detector resolution \cite{ic59}. The neutrino energy spectrum has to be inferred from the reconstructed muon energy.
This technique is denoted  as an inverse problem in unfolding, which is commonly defined by the Fredholm integral equation of the first kind \cite{fredholm03}.
In practice, an adequate algorithm is required to solve for the discrete problem set,
\begin{equation}
    \Vec{g}(y) = \textbf{A}(E_{\nu},y)\Vec{f}(E_{\nu}),
    \label{equ:discrete_fredholm}
\end{equation}
where the  spectral energy distribution  $\Vec{f}(E_{\nu})$  of the neutrinos has to be estimated from energy-dependent detector observables $y$. 
The response matrix $\textbf{A}(E_{\nu},y)$ accounts for propagation and detection effects and displays the smearing of the energy estimation. 

In this analysis, the neutrino energy spectrum is determined by the  Dortmund Spectrum Estimation Algorithm (DSEA)  \cite{dsea,ruheschmitz13,Ruhe16,leonora21}. The inverse problem is treated  as a multinomial classification task that is solved by an arbitrary supervised machine learning classifier. The classifier's predictions are accumulated for each pre-defined energy bin and the estimate is then updated iteratively to overcome potential biases. Supplementary  regularization methods within DSEA allow a scaling of the current estimate and thus accelerate the convergence behavior. 

The internal regularization parameters, the classifier selection and its internal settings are investigated in a tenfold cross-validation approach using simulated events from the IceCube neutrino-generator (NuGen) \cite{anis}. These events are generated with the assumption of an $E^{-2}$ power-law neutrino flux and are weighted to the atmospheric neutrino flux model Honda2006 \cite{honda06} to compensate for an over-representation of high energy events in the simulation sample. The optimization procedure yields the most accurate spectrum approximation measured by the Wasserstein Distance \cite{wasserstein_distance} with a default random forest classifier \cite{scikit-learn} in combination with exponential step-size decay regularization in seven iterations of  DSEA. 

\subsection{Analysis Scheme}

The seasonal spectra are obtained in ten logarithmic energy bins between \SI{125}{\giga\ev} and \SI{10}{\tera\ev}. 
Over- and underflow bins account for events outside of the respective interval. 
Multiple combinations of energy-dependent quantities are investigated in terms of acceptable coverage and minimal bias in \num{2000} trials of DSEA. The final selection consists of the number of hit DOMs, the number of direct hits, and the truncated neutrino energy: the number of direct hits is defined  assuming unscattered photons which arrive within a time residual from  $-\SI{15}{\nano\second}$ to \SI{75}{\nano\second}; the truncated energy is derived from the muon energy loss \cite{etrun13,ahrens02}. 

To account for statistical fluctuations in the unfolding, the spectral distributions are bootstrapped \cite{hastie2,ic79}. Each data set is sampled by replacement and deconvolved with optimized DSEA. The  final event spectra are determined by the average over 2000 deconvolutions, so that the standard deviation approximates the statistical uncertainty of the unfolding. Correcting for the effective detection area, livetime, and solid angle, the event spectra are transformed into a differential neutrino flux \cite{ic59}.



\subsection{Event Selection} 

To study the seasonal effects on the atmospheric neutrino flux, the diffuse upgoing event sample  which contains upward moving muons, as described in Ref.\ \cite{Aachen6yrs}, is divided into separate seasons. The analysis is developed using 10\% of the data taken between January 2011 and December 2018. Only events from the Southern Hemisphere in the  zenith range from \SIrange{90}{120}{\degree} are selected, excluding tropical latitudes between \SIrange{120}{135}{\degree}. This results in a total of \num{35038} events within 278 days of livetime, corresponding to approx.\ \num{3000} events per month.
 
\subsection{Systematic Uncertainties}

Since DSEA is trained on simulated events, the unfolded energy spectra are affected by systematic uncertainties in the detector simulation. The impact of these effects is estimated from simulations with varied parameters, similar to the approaches presented in \cite{ic59,antares21}. These generated events are treated as pseudo-data and are sampled according to the atmospheric neutrino flux model Honda2006 \cite{honda06}. The ratio of the unfolded spectrum to the reference unfolding with the default systematic parameters is the systematic uncertainty of the  parameter variation. Each uncertainty is combined in the squared sum for positive and negative deviations to the reference result and the total systematic uncertainty is given by, 
\begin{equation}
    \sigma_{\mathrm{sys}} = \sqrt{\sigma^2_{\mathrm{DOM}} + \sigma^2_{\mathrm{abs}} +  \sigma^2_{\mathrm{scat}} + \sigma^2_{\mathrm{flux}} },
    \label{equ:systematic_err}
\end{equation}
with the uncertainty caused by the efficiency of the optical modules $\sigma_{\mathrm{DOM}}$, the absorption and scattering coefficients of the ice model  $\sigma_{\mathrm{abs}}$ and $\sigma_{\mathrm{scat}}$, and the atmospheric neutrino flux model $\sigma_{\mathrm{flux}}$.

Monte Carlo  simulations from the same NuGen sample are used for the first three systematic parameters. The DOM efficiency was scaled by $ \pm 10\% $ \cite{Aartsen:2016}. 
Propagation effects of photons in the  ice are described by absorption and scattering coefficients. The simulated events are obtained from the \textit{SpiceLea} ice model, 
which accounts for the depth-dependence of both coefficients \cite{Aartsen:ice} and  anisotropies in the $xy$-plane of the detection volume \cite{chirkin13}. The dependence of both coefficients on one another are taken into account by a joint reduction of $-7\%$ as a lower bound for absorption and scattering. The upper bound is then determined for each parameter individually, increasing each by $+10\%$. The uncertainties induced by the ice model coefficients can be combined by Eq.\ (\ref{equ:systematic_err}) into a total  uncertainty of photon propagation effects.

Since the simulated events are weighted to a flux model for the algorithm optimization and  determination of the systematic uncertainties, the flux model imposes an additional uncertainty on the unfolded spectra. To estimate the impact of flux model uncertainties on the unfolded spectrum, the reference simulation is weighted according to the lower and upper flux limits. As illustrated in Ref.\ \cite{honda06}, the flux uncertainty scales linearly from \SI{100}{\giga\ev} to \SI{1}{\tera\ev} assuming an uncertainty of $14\%$ at \SI{100}{\giga\ev} and $25\%$ at \SI{1}{\tera\ev}. The uncertainty remains approximately constant  at energies between \SIrange{1}{10}{\tera\ev}.

To obtain an absolute measurement of seasonal neutrino fluxes, other sources of systematic uncertainties, such as assumptions about the primary cosmic ray spectrum, would have to be considered. However, since the central aim is to determine the seasonal variations with respect to the annual mean of the neutrino flux, the model independence of deconvolution is exploited and the variations are measured relatively to the annual mean. Assuming that the investigated detector systematics remain constant throughout the year, merely the propagation of statistical uncertainties impacts the flux deviation to the yearly average.


\section{Results}

\begin{figure}[htp]
    \subfigure[Austral summer and winter]{\includegraphics[width=0.49\textwidth]{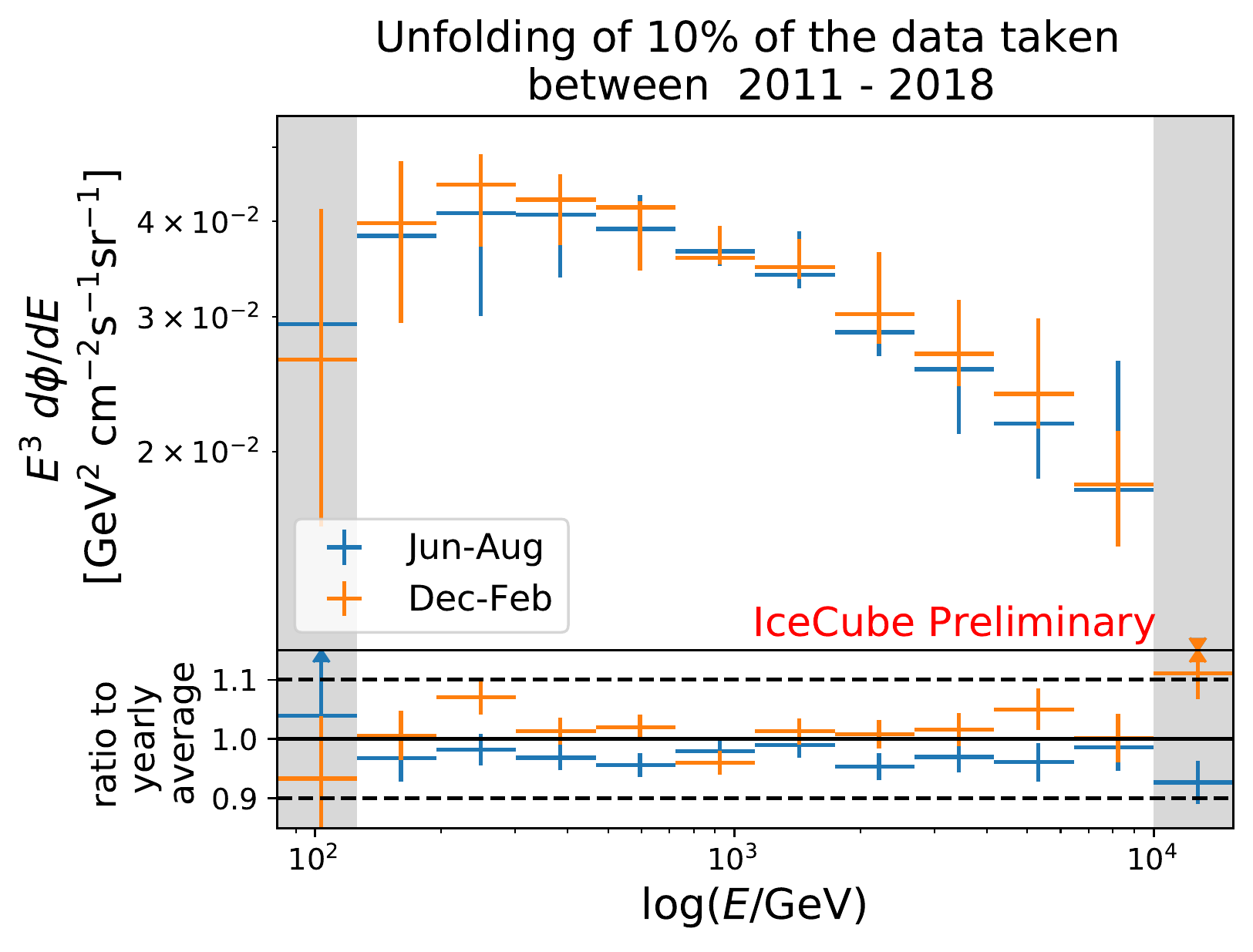}\label{fig:bns_result1}}
    \subfigure[Austral spring and autumn]{\includegraphics[width=0.49\textwidth]{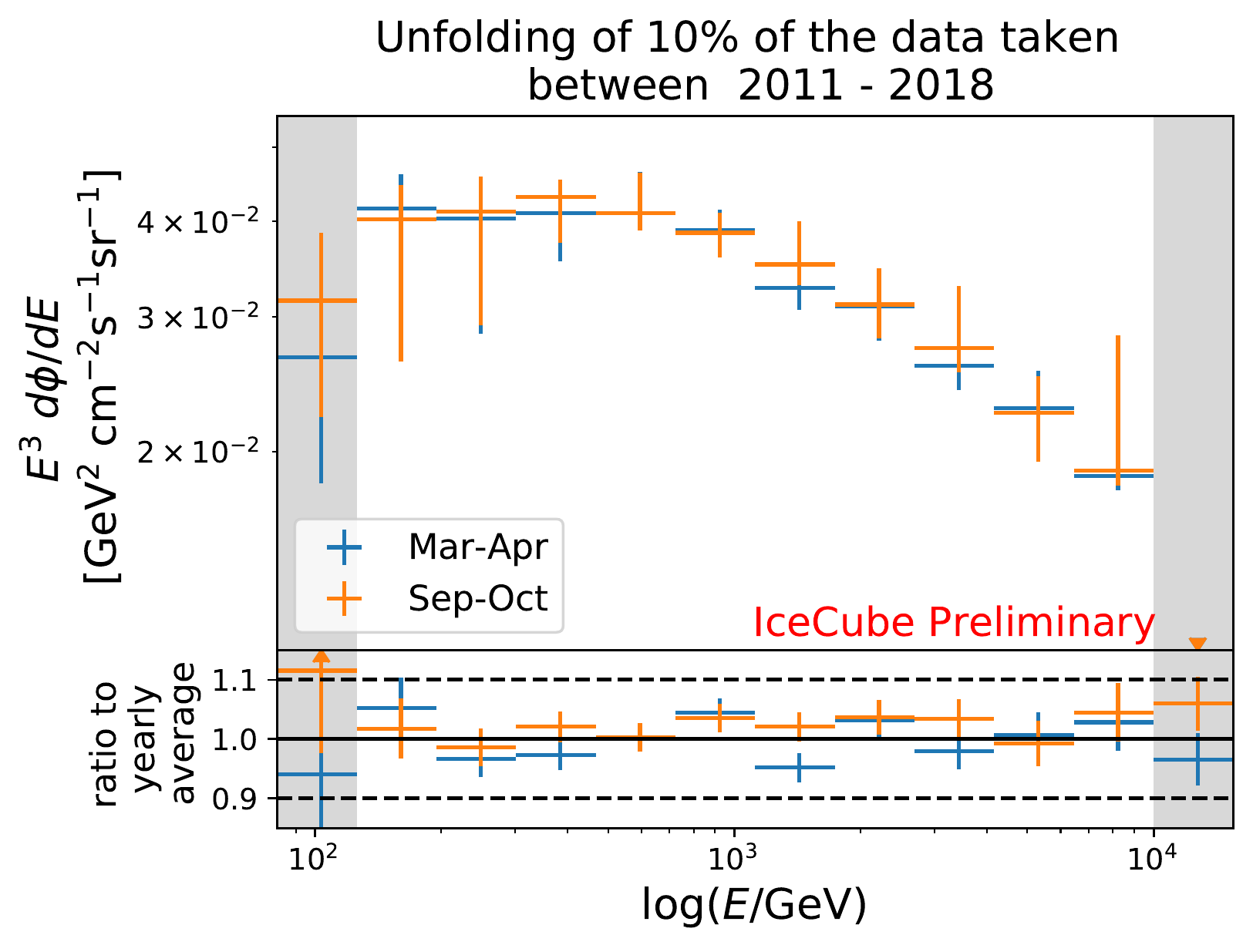}\label{fig:bns_result2}}
\caption{Unfolded seasonal muon neutrino spectra with statistical and systematic uncertainties using 10\% of the muon neutrino data from IceCube between 2011 and 2018 for the zenith range between \SIrange{90}{120}{\degree}. The  ratio to the unfolded flux averaged over all seasons is displayed below neglecting systematic uncertainties. First tendencies towards an increased flux in the Austral summer from December to February is observable despite large statistical uncertainties. The spectra for autumn and spring, on the other hand, are in agreement with the annual mean.} 

\end{figure}

The unfolded energy spectra for  Austral summer and winter are displayed in Fig.\ \ref{fig:bns_result1}. Both spectra agree within the uncertainties. Comparing the seasonal fluxes to the annual mean, a tendency towards an increased flux for the period from December to February is observable. The flux for the period from June to August remains smaller accordingly. A deviation around \SI{1}{\tera\ev} might be due to artifacts in the training sample. This was investigated in deconvolutions of monthly data sets.

The unfolded spectra for the Austral spring and autumn are displayed in Fig.\ \ref{fig:bns_result2} as a test of the unfolding method. Since both seasons should have similar temperature profiles, both spectra are expected to be in agreement. The data sets are constructed from events within two months from March to April and September to October to provide a clear demarcation between seasons. The seasonal energy spectra agree within the uncertainties and with the annual mean unfolded neutrino flux.

\begin{figure}[htp]
\centering
   \includegraphics[width=0.7\textwidth]{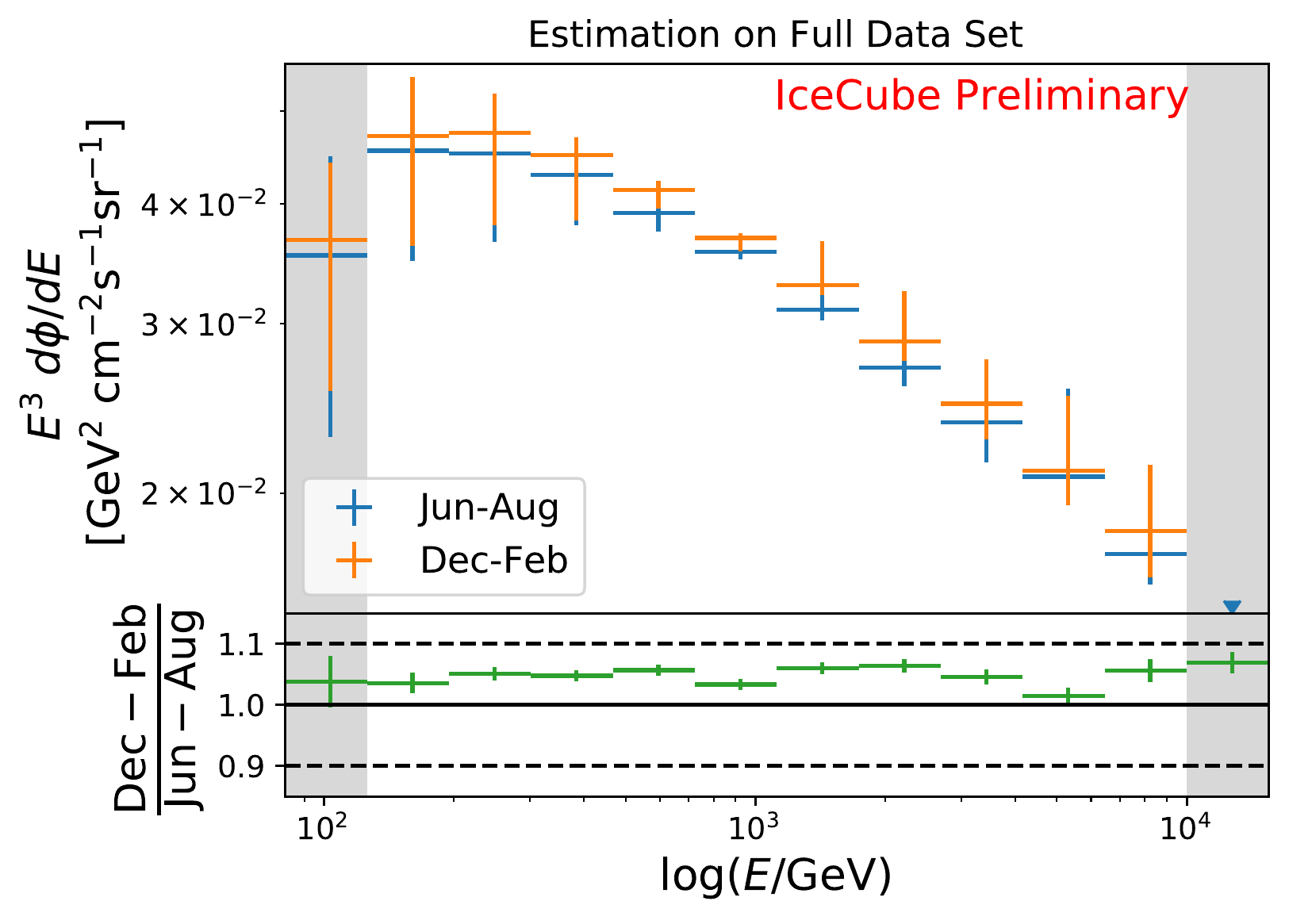}
\caption{Estimation of the unfolded seasonal spectra for the Austral summer and winter using all events from 2011 to 2018 as pseudo-data. Two sets are constructed by the Monte Carlo simulation weighted to the seasonal prediction from Ref.\  \cite{honda15}; scaling the number of events and livetime by a factor of ten. Both sets are then unfolded with DSEA and scaled to a differential flux. The ratio between the seasonal unfolded fluxes is displayed below in terms of statistical uncertainties.\label{fig:mc_estimation}}

\end{figure}

After the discussion of the unfolded spectra obtained from 10\% of the data taken from 2011 to 2018, the detection potential of seasonal variations using the complete data set is investigated on Monte Carlo simulations. For this, the simulated neutrino events are sampled  according to the seasonal flux model presented in Ref.\ \cite{honda15}. The number of events and the livetime of both pseudo-data sets were increased by a factor of ten.  The deconvolved spectra are denoted in Fig.\ \ref{fig:mc_estimation} with the associated uncertainties. Compared to the tests on 10\% of the data taken between 2011 and 2018, the uncertainties decrease significantly and the relationship of the seasonal neutrino fluxes to each other can be clearly observed below. This estimation yields a deviation of approx.\ 2\% to 8\% between the Austral summer and winter.

\section{Conclusion and Outlook}

The presented analysis holds a great potential for detecting seasonal variations of the atmospheric neutrino energy spectrum with IceCube data. Initial tendencies towards an increased flux in the Austral summer are evident on 10\% of the present atmospheric muon neutrino data set including events from 2011 to 2018. Due to the relative comparison of seasonal energy spectra, the analysis becomes independent of systematic uncertainties in the detector simulation and other model assumptions. An extension by a factor of ten to the entire data set would decrease the statistical uncertainty by the square root of ten, making seasonal variations to the unfolded annual mean flux more apparent. The sample can be extended to 9 years of data, further decreasing the statistical uncertainty, in particular at higher energies in the \si{\tera\ev}-regime. 

Further improvements of this analysis are in progress. The significance of the of the measured variations will be determined on 10\% of the recorded data. The hypothesis of observing no seasonal variations with respect to the annual mean will be investigated for each energy bin and season. Using the full 9-year data sample will potentially allow measurements of the monthly neutrino spectra with sufficient statistics for the first time.





\bibliographystyle{ICRC}
\bibliography{references}

\providecommand{\href}[2]{#2}\begingroup\raggedright\begin{thebibliography}{10}

\bibitem{desiati11}
{\bfseries IceCube} Collaboration
  \href{http://dx.doi.org/10.7529/ICRC2011/V01/0662}{{\em PoS} {\bfseries
  ICRC2011} (2012) 662}.

\bibitem{Gaisser13}
{\bfseries IceCube} Collaboration {\em PoS} {\bfseries ICRC2013} (2014) 0492.

\bibitem{Heix19}
{\bfseries IceCube} Collaboration {\em PoS} {\bfseries ICRC2019} (2020) 465.

\bibitem{honda15}
M.~Honda, M.~Sajjad~Athar, T.~Kajita, K.~Kasahara, and S.~Midorikawa
  \href{http://dx.doi.org/10.1103/PhysRevD.92.023004}{{\em Phys. Rev. D}
  {\bfseries 92} (2015) 023004}.

\bibitem{Aartsen:2016}
{\bfseries IceCube} Collaboration, M.~G. Aartsen {\em et~al.}
  \href{http://dx.doi.org/10.1088/1748-0221/12/03/P03012}{{\em JINST}
  {\bfseries 12} no.~03, (2017) P03012}.

\bibitem{gaisser90}
T.~K. Gaisser, {\em Cosmic rays and particle physics}.
\newblock Cambridge University Press, 1990.

\bibitem{nrlm}
J.~Picone, A.~Hedin, D.~Drob, and A.~Aikin
  \href{http://dx.doi.org/10.1029/2002JA009430}{{\em Journal of Geophysical
  Research} {\bfseries 107} (12, 2002) }.

\bibitem{ic59}
{\bfseries IceCube} Collaboration, M.~G. Aartsen {\em et~al.}
  \href{http://dx.doi.org/10.1140/epjc/s10052-015-3330-z}{{\em Eur. Phys. J. C}
  {\bfseries 75} no.~3, (2015) 116}.

\bibitem{etrun13}
{\bfseries IceCube} Collaboration, R.~Abbasi {\em et~al.}
  \href{http://dx.doi.org/10.1016/j.nima.2012.11.081}{{\em Nucl. Instrum. Meth.
  A} {\bfseries 703} (2013) 190--198}.

\bibitem{fredholm03}
I.~Fredholm \href{http://dx.doi.org/doi:10.1007/BF02421317}{{\em Acta Math.}
  {\bfseries 27} (1903) 365}.

\bibitem{dsea}
M.~Bunse {\em https://sfb876.tu-dortmund.de/deconvolution/index.html} .

\bibitem{ruheschmitz13}
T.~Ruhe, M.~Schmitz, T.~Voigt, and M.~Wornowizki {\em PoS} {\bfseries ICRC2013}
  (2014) 3354.

\bibitem{Ruhe16}
T.~Ruhe, T.~Voigt, M.~Wornowizki, M.~B{\"o}rner, W.~Rhode, and K.~Morik {\em
  Astronomical Data Analysis Software and Systems XXVI} {\bfseries 521} (2019)
  .

\bibitem{leonora21}
L.~Kardum {\em PoS} {\bfseries ICRC2021} (these proceedings) xyz.

\bibitem{anis}
A.~Gazizov and M.~P. Kowalski
  \href{http://dx.doi.org/10.1016/j.cpc.2005.03.113}{{\em Comput. Phys.
  Commun.} {\bfseries 172} (2005) 203}.

\bibitem{honda06}
M.~Honda, T.~Kajita, K.~Kasahara, S.~Midorikawa, and T.~Sanuki
  \href{http://dx.doi.org/10.1103/PhysRevD.75.043006}{{\em Phys. Rev. D}
  {\bfseries 75} (2007) 043006}.

\bibitem{wasserstein_distance}
A.~Ramdas, N.~Garcia, and M.~Cuturi
  \href{http://dx.doi.org/10.3390/e19020047}{{\em Entropy} {\bfseries 19}
  (2015) }.

\bibitem{scikit-learn}
F.~Pedregosa, G.~Varoquaux, A.~Gramfort, V.~Michel, B.~Thirion, O.~Grisel,
  M.~Blondel, P.~Prettenhofer, R.~Weiss, V.~Dubourg, J.~Vanderplas, A.~Passos,
  D.~Cournapeau, M.~Brucher, M.~Perrot, and E.~Duchesnay {\em Journal of
  Machine Learning Research} {\bfseries 12} (2011) 2825--2830.

\bibitem{ahrens02}
{\bfseries AMANDA} Collaboration, J.~Ahrens {\em et~al.}
  \href{http://dx.doi.org/10.1103/PhysRevD.66.012005}{{\em Phys. Rev. D}
  {\bfseries 66} (2002) 012005}.

\bibitem{hastie2}
T.~Hastie, J.~Friedman, and R.~Tibshirani, {\em The Elements of Statistical
  Learning: Data Mining, Inference, and Prediction}.
\newblock Springer Publishing Company, Incorporated, 2017.

\bibitem{ic79}
{\bfseries IceCube} Collaboration, M.~G. Aartsen {\em et~al.}
  \href{http://dx.doi.org/10.1140/epjc/s10052-017-5261-3}{{\em Eur. Phys. J. C}
  {\bfseries 77} no.~10, (2017) 692}.

\bibitem{Aachen6yrs}
{\bfseries IceCube} Collaboration, M.~G. Aartsen {\em et~al.}
  \href{http://dx.doi.org/10.3847/0004-637X/833/1/3}{{\em ApJ} {\bfseries 833}
  (2016) 3}.

\bibitem{antares21}
{\bfseries ANTARES} Collaboration, A.~Albert {\em et~al.}
  \href{http://dx.doi.org/10.1016/j.physletb.2021.136228}{{\em Phys. Lett. B}
  {\bfseries 816} (2021) 136228}.

\bibitem{Aartsen:ice}
{\bfseries IceCube} Collaboration, M.~Aartsen {\em et~al.}
  \href{http://dx.doi.org/10.1016/j.nima.2013.01.054}{{\em Nucl. Instrum. Meth.
  A} {\bfseries 711} (2013) 73--89}.

\bibitem{chirkin13}
{\bfseries IceCube} Collaboration {\em PoS} {\bfseries ICRC2013} (2014) 0580.

\end{thebibliography}\endgroup



\clearpage
\section*{Full Author List: IceCube Collaboration}




\scriptsize
\noindent
R. Abbasi$^{17}$,
M. Ackermann$^{59}$,
J. Adams$^{18}$,
J. A. Aguilar$^{12}$,
M. Ahlers$^{22}$,
M. Ahrens$^{50}$,
C. Alispach$^{28}$,
A. A. Alves Jr.$^{31}$,
N. M. Amin$^{42}$,
R. An$^{14}$
% ...
\\

\noindent
% ...
$^{12}$ Universit{\'e} Libre de Bruxelles, Science Faculty CP230, B-1050 Brussels, Belgium \\
% ...
$^{14}$ Department of Physics and Laboratory for Particle Physics and Cosmology, Harvard University, Cambridge, MA 02138, USA \\
% ...
$^{17}$ Department of Physics, Loyola University Chicago, Chicago, IL 60660, USA \\
$^{18}$ Dept. of Physics and Astronomy, University of Canterbury, Private Bag 4800, Christchurch, New Zealand \\
% ...
$^{22}$ Niels Bohr Institute, University of Copenhagen, DK-2100 Copenhagen, Denmark \\
% ...
$^{28}$ D{\'e}partement de physique nucl{\'e}aire et corpusculaire, Universit{\'e} de Gen{\`e}ve, CH-1211 Gen{\`e}ve, Switzerland \\
% ...
$^{31}$ Karlsruhe Institute of Technology, Institute for Astroparticle Physics, D-76021 Karlsruhe, Germany  \\
% ...
$^{42}$ Bartol Research Institute and Dept. of Physics and Astronomy, University of Delaware, Newark, DE 19716, USA \\
% ...
$^{50}$ Oskar Klein Centre and Dept. of Physics, Stockholm University, SE-10691 Stockholm, Sweden \\
% ...
$^{59}$ DESY, D-15738 Zeuthen, Germany \\
% ...

\subsection*{Acknowledgements}

\noindent
USA {\textendash} U.S. National Science Foundation-Office of Polar Programs,
U.S. National Science Foundation-Physics Division,
U.S. National Science Foundation-EPSCoR,
Wisconsin Alumni Research Foundation,
Center for High Throughput Computing (CHTC) at the University of Wisconsin{\textendash}Madison,
Open Science Grid (OSG),
Extreme Science and Engineering Discovery Environment (XSEDE),
Frontera computing project at the Texas Advanced Computing Center,
U.S. Department of Energy-National Energy Research Scientific Computing Center,
Particle astrophysics research computing center at the University of Maryland,
Institute for Cyber-Enabled Research at Michigan State University,
and Astroparticle physics computational facility at Marquette University;
Belgium {\textendash} Funds for Scientific Research (FRS-FNRS and FWO),
FWO Odysseus and Big Science programmes,
and Belgian Federal Science Policy Office (Belspo);
Germany {\textendash} Bundesministerium f{\"u}r Bildung und Forschung (BMBF),
Deutsche Forschungsgemeinschaft (DFG),
Helmholtz Alliance for Astroparticle Physics (HAP),
Initiative and Networking Fund of the Helmholtz Association,
Deutsches Elektronen Synchrotron (DESY),
and High Performance Computing cluster of the RWTH Aachen;
Sweden {\textendash} Swedish Research Council,
Swedish Polar Research Secretariat,
Swedish National Infrastructure for Computing (SNIC),
and Knut and Alice Wallenberg Foundation;
Australia {\textendash} Australian Research Council;
Canada {\textendash} Natural Sciences and Engineering Research Council of Canada,
Calcul Qu{\'e}bec, Compute Ontario, Canada Foundation for Innovation, WestGrid, and Compute Canada;
Denmark {\textendash} Villum Fonden and Carlsberg Foundation;
New Zealand {\textendash} Marsden Fund;
Japan {\textendash} Japan Society for Promotion of Science (JSPS)
and Institute for Global Prominent Research (IGPR) of Chiba University;
Korea {\textendash} National Research Foundation of Korea (NRF);
Switzerland {\textendash} Swiss National Science Foundation (SNSF);
United Kingdom {\textendash} Department of Physics, University of Oxford.

\end{document}